\documentclass[12pt, leqno]{article}
\usepackage{graphicx}
\usepackage[latin1]{inputenc}
\usepackage{amsmath}

\begin{document}

\title{Monte Carlo simulations of the Ising and the Sznajd model
on growing Barabási - Albert networks}

\author{\large{Johannes Bonnekoh}\footnote{Present address: Mathematical Institut, University
of Cologne, Weyertal 86 - 90, D-50931 Köln, Germany} \\ Institute
of Theoretical
Physics, University of Cologne, \\ D-50923 Köln, Germany \\
E-Mail: jbonneko@mi.uni-koeln.de}

\date{}

\maketitle

\section*{Abstract}

The Ising model shows on growing Barabási - Albert networks the
same ferromagnetic behavior as on static Barabási - Albert
networks. Sznajd models on growing Barabási - Albert networks show
an hysteresis like behavior. Nearly a full consensus builds up and
the winning opinion depends on history. On slow growing Barabási -
Albert networks a full consensus builds up. At five opinions in the
Sznajd model with limited persuasion on growing Barabási - Albert
networks, all odd opinions win and all even opinions loose supporters.

\bigskip

\textbf{Keywords:} Monte Carlo simulations, Sociophysics, Ising
model, Sznajd model, limited persuasion, growing Barabási - Albert networks

\section{The Ising model}

First, we initiate a Barabási - Albert network \cite{ba} with $m$
nodes. Each node is connected to every other node with exactly one
connection. At every time step we put one new node to the network.
This new node builds up randomly $m$ connections to already
existing nodes. The probability for a existing node to be chosen
as a neighbor is proportional to the number of its neighbors.
After putting a new node on the network, we go through the whole
network and use the Ising model on every node. We say that we make
one Ising rpts (run per time step).

\begin{figure}
\includegraphics [angle=-90,scale=0.5]{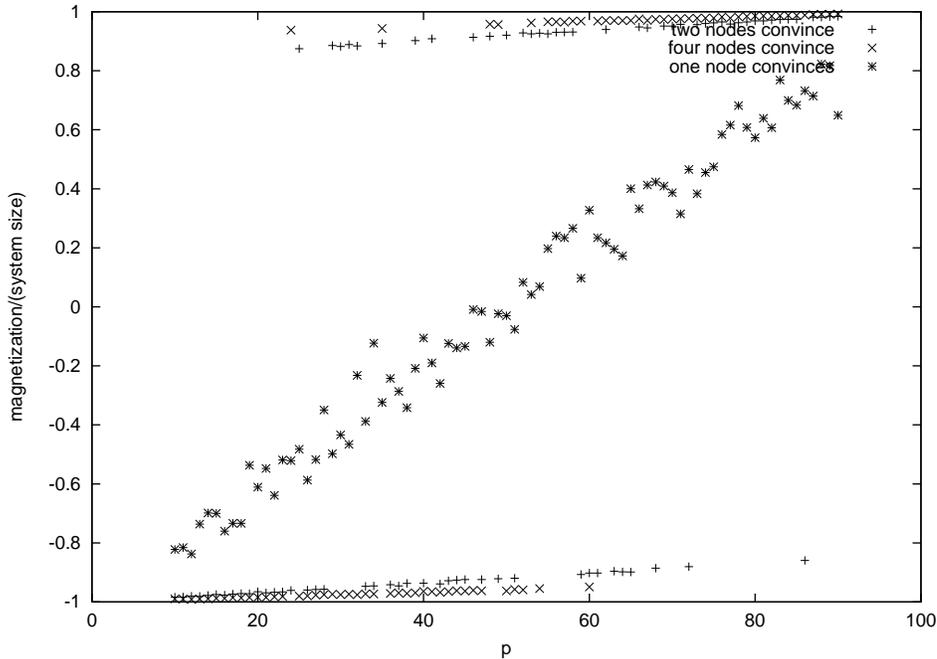}
\caption{\label{s4sweepnsweep} Sznajd model on a 25000 nodes
network with one Sznajd run per time step, $m=4$.}
\end{figure}

The explanation of the classical Ising model and how to simulate its Glauber
kinetics can be found in \cite{binder}.

The simulations show the same ferromagnetic behavior as given in
\cite{alek} for static Barabási - Albert networks. Even when we
make more than one Ising rpts the results do not change.

\section{The Sznajd model}

We initiate the network as the Ising model. At every time
step we put one new node to the network. After that we randomly
chose one node and a few (none, one, or three, depending on the model used)
of its neighbors. If all these randomly
chosen nodes have the same opinion (spin), they convince all their
neighbors \cite{sznajd}. We are using the opinions $+1$ and $-1$.

Initially, every node has with probability $0\le p\le 100$
the opinion $+1$ and with probability $1-p$ the opinion $-1$.

For the number of randomly chosen nodes we have a few
possibilities:

\begin{figure}
\begin{center}
\includegraphics [angle=-90,scale=0.5]{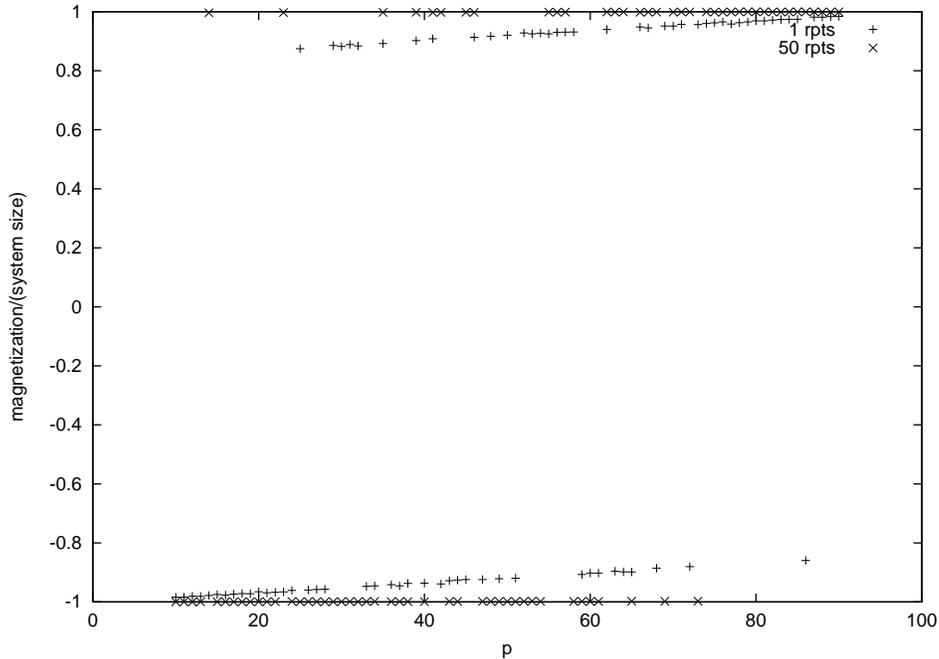}
\end{center}
\caption{\label{s4sweepnrunsweep}Sznajd model on a 25000 nodes
network with $m=4$ and two nodes convincing}
\end{figure}

\begin{itemize}
\item two randomly chosen neighbors,
\item four randomly chosen neighbors or
\item one randomly chosen node
\end{itemize}
convince all their neighbors.

In Figure \ref{s4sweepnsweep} we can see that the two- and
four-nodes-convincing models come to the same results. In these
models nearly a consensus builds up. The one-node-convincing model
does not show the same behavior as the other models. We can see
too that in the two- or four-nodes-convincing model no sharp
opinion change takes place. We have a area of coexistence of both
opinions. Like in magnetic hysteresis, the final magnetization is
close to $-1$ or $+1$, depending on history (initial
configuration); averages over 100 samples
give an average value far away from
consensus ($+1$ or $-1$).

The results shown in Figure \ref{s4sweepnsweep} are nearly equal
for every value of $m$.

\begin{figure}
\begin{center}
\includegraphics [angle=-90,scale=0.5]{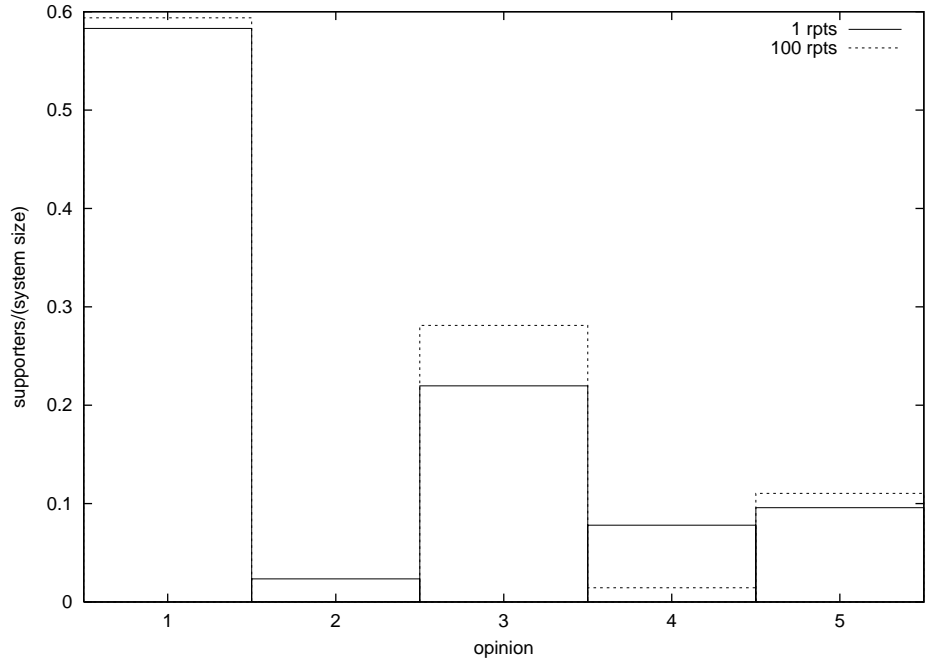}
\end{center}
\caption{\label{s4m4limitsweepnrunsweep}Sznajd model with limited
persuasion on a 25000 nodes network with $m=4$ and two nodes
convincing}
\end{figure}

If we make more Sznajd runs per time step we will see, that a
consensus builds up. This result is shown in Figure
\ref{s4sweepnrunsweep} for the two nodes convincing model. The
results are equal to the ones of the four nodes convincing model.
We can see too that the area of coexistence becomes smaller for
more Sznajd runs per time step.

\section{The Sznajd model with limited persuasion}

Now I simulated a system with the five opinions 1, 2, 3, 4 and 5.
Therefore I used the Sznajd model with limited persuasion \cite {stauffer}
on growing Barabási - Albert networks. A new node has with
probability $0\le p_i\le 100$ the opinion $i$. The initial probabilities

\begin{itemize}
\item $p_1=34$,
\item $p_2=29$,
\item $p_3=14$,
\item $p_4=13$ and
\item $p_5=10$
\end{itemize}
have been used.

The procedure is mainly the same as for the classical Sznajd model
but node $i$ can convince node $j$ only if \cite{deffuant,hegs}
$$\left|\text{opinion}\left(i\right)-\text{opinion}\left(i\right)\right|
=1.$$ For the simulations the two-nodes-convincing model has been
used.

The main results can be seen in Figure
\ref{s4m4limitsweepnrunsweep}. All even opinions loose supporters
while all odd opinions win supporters. Opinion $1$ wins the most
supporters and has on a $25000$ nodes network nearly $60$\%,
depending on the value of $m$ at one Sznajd run per time step.
Opinion 2 loosest the most supporters.

If we make more than one Sznajd run per time step the difference
becomes even stronger. At small values of $m$ opinion 2 is
nearly not existing.

\bigskip

I thank D. Stauffer for his support and help during
my work at these models.

\end{document}